\documentclass[fleqn]{annalen1}
\usepackage{latexsym}
\begin{document}
%%%%%%%%%%%%%%%%%%%%%%%%%%%%%%%%%%%%%%%%%%%%%%%%%%%%%%%%%%%%%%%%%%%%%%%%%%%%%%
\newcommand{\keywords}{anti de Sitter, Chern-Simon, Liouville} 
%%%%%%%%%%%%%%%%%%%%%%%%%%%%%%%%%%%%%%%%%%%%%%%%%%%%%%%%%%%%%%%%%%%%%%%%%%%%%%
\newcommand{\PACS}{11.10.Kk, 04.20.Ha}
%%%%%%%%%%%%%%%%%%%%%%%%%%%%%%%%%%%%%%%%%%%%%%%%%%%%%%%%%%%%%%%%%%%%%%%%%%%%%%
\newcommand{\shorttitle} {Aspects of (2+1) dimensional gravity} 
%%%%%%%%%%%%%%%%%%%%%%%%%%%%%%%%%%%%%%%%%%%%%%%%%%%%%%%%%%%%%%%%%%%%%%%%%%%%%%
\title{Aspects of (2+1) dimensional gravity: $\rm AdS_3$ asymptotic dynamics 
in the framework of Fefferman-Graham-Lee theorems}
%%%%%%%%%%%%%%%%%%%%%%%%%%%%%%%%%%%%%%%%%%%%%%%%%%%%%%%%%%%%%%%%%%%%%%%%%%%%%%
\author{M.\ Rooman$^{1}$, Ph.\ Spindel$^{2}$} 
%%%%%%%%%%%%%%%%%%%%%%%%%%%%%%%%%%%%%%%%%%%%%%%%%%%%%%%%%%%%%%%%%%%%%%%%%%%%%%
\newcommand{\address}
  {$^{1}$Service de Physique Th\'eorique, CP 225,
  Universit\'e Libre de Bruxelles, 1050 Bruxelles,
  \\ \hspace*{0.5mm} Belgium \\ 
  $^{2}$M\'ecanique et Gravitation, Universit\'e de Mons-Hainaut,
   7000 Mons,
   \\ \hspace*{0.5mm} Belgium}
%%%%%%%%%%%%%%%%%%%%%%%%%%%%%%%%%%%%%%%%%%%%%%%%%%%%%%%%%%%%%%%%%%%%%%%%%%%%%%
\newcommand{\email}{\tt mrooman@ulb.ac.be, spindel@umh.ac.be} 
\maketitle 
%%%%%%%%%%%%%%%%%%%%%%%%%%%%%%%%%%%%%%%%%%%%%%%%%%%%%%%%%%%%%%%%%%%%%%%%%%%%%
\newcommand{\real}{{\hbox{{\rm I}\kern-.2em\hbox{\rm R}}}}
\def\mettre#1\sous#2{\mathrel{\mathop{\kern0pt #2}\limits_{#1}}}

\begin{abstract}
Using the Chern-Simon formulation of (2+1) gravity, 
we derive, for the general asymptotic metrics given by the
Fefferman-Graham-Lee theorems, the emergence of the Liouville mode 
associated to the boundary degrees of freedom of (2+1) dimensional 
anti de Sitter geometries.
\end{abstract}
%%%%%%%%%%%%%%%%%%%%%%%%%%%%%%%%%%%%%%%%%%%%%%%%%%%%%%%%%%%%%%%%%%%%%%%%%%%%%
\section{Introduction}

The interest of studying (2+1) dimensional gravity has initially been 
emphasized in \cite{JD} and has recently been revived with the discovery 
of black holes in spaces with negative cosmological constant \cite {BTZ}.
Since then, a large number of studies has been devoted to the elucidation
of classical as well as to quantum (2+1) gravity. 

In particular, we examined \cite{LRS} 
stellar-like models corresponding to
stationary, rotationally symmetric gravitational sources of the perfect 
fluid type, embedded in spaces of arbitrary cosmological constant, and showed
how causality privileges anti de Sitter (AdS) backgrounds. As this part of the
talk has already been published, we will not re-describe it here.

On the other hand, $(2+1)$
gravity with negative cosmological constant has been proven to be equivalent 
to Chern-Simons 
(CS) theory with $SL(2,\real) \times SL(2,\real)$ as gauge group \cite{TW}.
Assuming the boundary of the space to be a flat cylinder $\real \times S^1$, 
Coussaert, Henneaux and van Driel (CHD) \cite{CHD} demonstrated 
the equivalence 
between this CS theory and a non-chiral Wess-Zumino-Witten (WZW)
theory \cite{WZW}, and showed that the $\rm AdS_3$ boundary conditions as defined 
in \cite{BH} implement the constraints
that reduce the WZM model to the Liouville theory \cite{FWB}.
 
In this short note, we show that, using the less restrictive
AdS boundary conditions deduced from the Fefferman-Graham-Lee
theorems \cite{GL,FG}, the CHD analysis can be extended and leads
to the Liouville theory formulated on a 2-dimensional curved background.
A detailed version of this work will be found in \cite {BERS}.

\section{Asymptotically anti de Sitter spaces}

Graham and Lee \cite{GL} proved that, under suitable topological
assumptions, Euclidean
Einstein spaces with negative cosmological constant $\Lambda$
are completely defined by the geometry on
their boundary. Furthermore, Fefferman and Graham \cite{FG} 
showed that, whatever the signature,
there exists a formal asymptotic expansion of the metric, which formally 
solves the Einstein equations with $\Lambda<0$.
The first terms of this expansion are given by even powers of 
a radial coordinate $r$:
\begin{equation}
\label{3}
ds^2 
\begin{array} {c} \\ \approx \\ 
{ {r \rightarrow \infty}} \end{array}
 \ell^2 {dr^2\over{r^2}} \, + \,
\frac {r^2} {\ell^2} \stackrel {(0)} {\bf g} (x^i)  \, + \,
\stackrel {(2)} {\bf g} (x^i) \, + \, \cdots \qquad .
\end{equation}
On $d$-dimensional space-times, the full asymptotic expansion actually 
continues with terms of negative even  powers of $r$ up
to $r^{-2([\frac d 2]-2)}$, with in addition a logarithmic term of the order
of $r^{-(d-3)}\log r$ when $d$ is odd and larger than 3. All these terms
are completely defined by the boundary geometry. They are
followed by terms of all negative powers
starting from $r^{-(d-3)}$; the trace-free part of the $r^{-(d-3)}$
coefficient is not fully determined by $\stackrel {(0)} {\bf g}$
but contains degrees of freedom \cite{GW}. 

It is instructive to look at the first iterations of this
expansion to see the special character of 3 dimensions. We therefore write
the metric in terms of forms $\underline{\Theta}^{ \mu}$ as
$ds^2 = \underline{\Theta}^{ 0} \otimes \underline{\Theta}^{ 0} +
\eta_{ a b} \,
\underline{\Theta}^{ a} \otimes \underline{\Theta}^{ b}$
with $\mu, \nu$ [resp. $a, b$] running from 0 to $n$ [resp. 1 to $n$] and $\eta _{ a b}$ a
flat $n$-dimensional minkowskian metric {\it diag.}(1,..., 1, -1). The forms 
$\underline{\Theta}^{ \mu}$ read as:
\begin{equation}
\label{5}
\underline{\Theta}^{ 0} = \ell{dr\over r} \quad , \quad
\underline{\Theta}^{ a} =
{r\over{\ell}} \underline{\theta}^{ a} + {\ell\over r}
\underline{\sigma}^{ a} +
O (r^{-3}) \qquad ,
\end{equation}
where the forms $\underline {\theta}^{ a}$ and 
$\underline {\sigma}^{ a} \equiv  {\sigma}^{ a}_{\  b} 
\, \underline{\theta}^{ b}$
are r-independent. These provide the dominant 
and sub-dominant terms of the metric expansion:
\begin{equation}
\label{6}
\stackrel {(0)}{\bf g}= \eta_{a   b}\, \underline{\theta}^{ a} \otimes
\underline{\theta}^{ b}
\quad , \quad 
\stackrel {(2)}{\bf g} \equiv
\stackrel {(2)} g _{a b} \underline{\theta}^{ a}\otimes
 \underline{\theta}^{ b}
= \left ( \sigma_{ a  b} +  \sigma_{ b  a}\right )
\underline{\theta}^{ a} \otimes \underline{\theta}^{ b} \qquad .
\end{equation}
Here and in what follows, the n-dimensional indices
and the covariant derivatives are defined
with respect to the metric $\stackrel {(0)}{\bf g}$.
Using these definitions, the components of the (n+1)-dimensional
Riemann curvature 2-form 
$\underline{\underline R}$ become:
\begin{equation}
\label{9}
\underline{\underline R}_{ a 0} =
- \frac 1 {\ell^2} \underline{\Theta} _{ a} \wedge \underline{\Theta}_{
0}-{1\over r}
\left(d\underline{\gamma}_{ a} + \underline{\omega}_{ a b}
\wedge \underline{\gamma}^{ b}
\right) + O(r^{-3}) \qquad ,
\end{equation}
\begin{equation}
\label{10}
\underline{\underline R}_{ a b} =
 - \frac 1 {\ell^2} \underline{\Theta}_{ a} \wedge
\underline{\Theta}_{ b}+
\stackrel {(0)} {\underline{\underline {\cal R}}} _{ a b} +
\frac 1 {\ell^2} (\underline{\theta}_{ a}\wedge \underline{\gamma}_{ b}
+ \underline{\gamma}_{ a} \wedge \underline{\theta}_{ b}) + O(r^{-2}) \qquad ,
\end{equation}
where ${\underline \omega}_{a b}$ is the n-dimensional Levi-Civita connection
and
$ \stackrel {(0)} {\underline{\underline {\cal R}}} _{ a b}$ 
the n-dimensional curvature 2-form, both 
defined by the metric $\stackrel {(0)}{\bf g}$,
and $\underline{\gamma}_{ a}
\equiv \stackrel {(2)} g _{ a  b}\underline{\theta}^{ b}$.
If we impose the metric of the (n+1)-dimensional space
to be Einsteinian, these equations, at order $r^2$, fix $\Lambda = -1/\ell^2$. 
Moreover, at order 1 and $r^{-1}$, they yield:
\begin{equation}
\label{13}
\stackrel {(0)} {\cal{R}}_{ a  b} + {1\over{\ell^2}} [(n-2)
\stackrel {(2)}g_{  a   b} + 
\eta _{  a   b} \stackrel {(2)}g \strut _{ c}^{ c}] = 0 \qquad ,
\end{equation}
\begin{equation}
\label{12}
\stackrel {(2)} g \strut  ^{ b}_{ b;  a}- 
\stackrel {(2)} g \strut ^{ b}_{ a;  b} = 0 \qquad ,
\end{equation}
where $\stackrel {(0)}{\cal R}_{ a  b}$ are the components of the $n$-dimensional Ricci 
tensor.
These equations clearly reveal the pecularity of 3-dimensional spaces.
Indeed, when $n \neq 2$,
eq. (\ref{13}) fully specifies the metric $\stackrel {(2)}{\bf g}$ and 
eq. (\ref{12})
becomes the Bianchi identity satisfied by the $n$-dimensional Einstein 
tensor. If we moreover require the space to be asymptotically AdS, 
the finite terms in eqs (\ref{9}-\ref{10}) have to vanish, 
thereby implying the $n$-dimensional geometry to be conformally flat.

On the contrary, when $n=2$, only the trace of $\stackrel {(2)}{\bf g}$ 
is fixed by eq. (\ref{13}):
\begin{equation}
\label{14}
\stackrel {(2)}{g} \strut ^{ c}_{  c}= 2\,
\sigma^{  c}_{  c} \equiv 2 \, \sigma = -{\ell^2\over 2}\stackrel {(0)}{\cal R}
\qquad ,
\end{equation}
and the other components of $\stackrel {(2)}{\bf g}$ have only
to satisfy the equations:
\begin{equation}
\label{15}
\stackrel {(2)}g \strut ^{ a}_{ b \, ;  a} = -{\ell^2\over 2} 
\stackrel {(0)} {\cal R} _{, b }
\qquad .
\end{equation}
The subdominant metric components are thus not all determined by 
the asymptotic metric in 3 dimensions, but there remains one
degree of freedom, which we shall explicit in the next section.
Note that in 3 dimensions Einstein spaces with $\Lambda <0$ are 
locally AdS and metrics on cylindrical boundaries are conformally flat;
this implies the equivalence between eqs (\ref{13},\ref{12}) and the
vanishing of the sub-dominant terms on the right-hand side of eqs
(\ref{9},\ref{10}).

\section{From Einstein-Hilbert to Liouville action}

The Einstein-Hilbert (EH) (2+1) gravity action with $\Lambda <0$
is equivalent to the difference of two CS
actions $S_{CS}[{\bf A}] - S_{CS}[{\bf \tilde A}]$ with
\begin{equation}
\label{17}
S_{CS}[{\bf A}] = {1\over 2} \int Tr ({\bf A}\wedge d{\bf A} + {2\over 3} {\bf
  A}\wedge {\bf A} \wedge {\bf A}) \qquad .
\end{equation}
The gauge fields 
${\bf A}=A_\mu {\underline \Theta} ^\mu = J_{ \mu} \, \underline {A}^{ \mu}$, 
with $J_{ \mu}$
generators of the $sl(2,\real)$ algebra\footnote{
We use the conventions:
$J_0=\frac 1 2 \left ( \begin{array} {cc} 1 & 0 \\ 0 & -1  \end{array} 
\right )$,
$J_1=\frac 1 2 \left ( \begin{array} {cc} 0 & 1 \\ 1 & 0  \end{array} 
\right )$ and  
$J_2=\frac 1 2 \left ( \begin{array} {cc} 0 & -1 \\ 1 & 0  \end{array} 
\right )$.},
are given as functions of the 3-bein
$\underline{\Theta}^{ \mu}$ and the Levi-Civita connection form
$\underline{\Omega}^{ \mu  \nu}$ by:
\begin{equation}
\label{19}
\underline{A}^{ \mu} = \frac 1 \ell \underline{\Theta}^{ \mu} +
{1\over 2} \epsilon^{ \mu}\ _{
\nu\rho} \underline{\Omega}^{ \nu  \rho}  \qquad , \qquad
\underline{\tilde A}^{ \mu} = - \frac 1 \ell \underline{\Theta}^{ \mu} +
{1\over 2} \epsilon^{ \mu}\ _{
\nu\rho} \underline{\Omega}^{ \nu  \rho} \qquad (\epsilon _{012}=1) .
\end{equation}
In cylindrical coordinates
$\lbrace r, \phi, t \rbrace$, the CS action may be written as:
\begin{eqnarray}
\label{21}
S_{CS}[{\bf A}] &=& \frac 1 2 \int_{\cal M}{ Tr (2\ A_t F_{r\phi} + 
A_{\phi}\dot A_r -A_r \dot A_{\phi} )}
 \, dr \, d\phi \, dt \, + \, S_B [{\bf A}] \qquad, \\
S_B [{\bf A}] &=& - {1\over 2} \int_{\partial \cal M} Tr (A_t A_\phi) \, d\phi\, dt
\qquad,
\end{eqnarray}
where $\cal M$ stands for bulk. The on-shell variation of this action 
is given by:
\begin{equation}
\label{22}
\delta S_{CS}[{\bf A}] 
=  \frac 1 2 \int_{\partial \cal M} 
Tr ( A_t \delta A_{\phi} -  A_\phi \delta A_{t} )
\, d\phi\, dt \qquad .
\end{equation}
The asymptotic behaviour of the fields ${\bf A}$ and ${\bf \tilde A}$,
dictated by the ${\rm AdS_3}$ boundary conditions, are easily expressed
using the null frame ${\underline \theta}^{\pm} = {\underline \theta}^{1} 
\pm {\underline \theta}^{2}$ and its 
dual vectorial frame ${\vec e}_{\pm}= \frac 1 2 ({\vec e}_{1} 
\pm {\vec e}_{2})$. Indeed, the null components $A_{-}$ and 
$\tilde A_{+}$ do not contain any degrees of freedom, i.e. they only
depend on the metric $\stackrel {(0)} {\bf g}$ and on 
its scalar curvature (see eq. \ref{14}).
At order $r^{-1}$, they are equal to:
\begin{equation}
\label{BC}
A_{-} = \frac 1 2 \left(\begin{array}{cc} 
 \omega_{-} &  \frac \sigma r \\
  0         &  - \omega_{-} 
  \end{array}\right) \equiv K_{-} \quad , \quad
\tilde A_{+} = \frac 1 2 \left(\begin{array}{cc} 
 \omega_{+} &  0 \\
  - \frac \sigma r &  - \omega_{+} 
  \end{array}\right) \equiv \tilde K_{+} \qquad ,
\end{equation}
where we have introduced the null components of the connection 2-form
${\underline \omega}^{\phi t} = \omega_{+} {\underline \theta} ^{+} +
\omega_{-} {\underline \theta} ^{-}$. The other components:
\begin{equation}
\label{BCBC}
A_{+} = \left(\begin{array}{cc} 
 \frac {\omega_{+}} 2  &  \frac  {\sigma^{-}_{\ +}} r \\
  \frac {r} {\ell^2} + \frac {\sigma^{+}_{\ +}-\sigma^{-}_{\ -}}{2r} 
   &  - \frac {\omega_{+}} 2
  \end{array}\right) \quad , \quad
\tilde A_{-} = \left(\begin{array}{cc} 
 \frac {\omega_{-}} 2 & 
 \frac {-r} {\ell^2} + \frac {\sigma^{+}_{\ +}-\sigma^{-}_{\ -}}{2r}  \\
    \frac  {- \sigma^{+}_{\ -}} r  &  -  \frac {\omega_{-}} 2
  \end{array}\right)  \quad ,
\end{equation}
depend on the dynamical part of $\stackrel {(2)} {\bf g}$, which is not
fixed by $\stackrel {(0)} {\bf g}$. This implies that 
$\delta A_{-}=O(r^{-2})=\delta \tilde A_{+}$ and
$A_{-} \delta A_{+}=O(r^{-2})= \tilde A_{+} \delta \tilde A_{-}$.
So, rewriting the variation of the action (\ref{22}) in 
terms of null components yields:
\begin{equation}
\delta S_{CS}[{\bf A}] 
=  \frac 1 2 \int_{\partial \cal M} 
Tr ( A_{+} \delta A_{-}  -  A_{-} \delta A_{+} )
\, \theta \, d\phi\, dt  = 
\int_{\partial \cal M} O(r^{-2}) \, d\phi\, dt \qquad ,
\end{equation}
with $\theta= \theta _t^{+} \theta _\phi^{-} -\theta _t^{-} \theta _\phi^{+}$.
It is thus easy to see that, owing to the boundary conditions 
(\ref{BC},\ref{BCBC}), 
the variation of the action $S_{SC}$ vanishes, 
without the addition of any extra boundary term. However, as the practical 
implementation of the boundary condition (\ref{BCBC}) is not obvious
at this stage, we prefer to modify the action by the addition of 
the boundary term
\begin{equation}
S^\prime _B [{\bf A}]
=  \frac 1 2 \int_{\partial \cal M}  
Tr (  A_{-} \,  A_{+}) \, \theta \, d\phi\, dt  \qquad ,
\end{equation}
which ensures that $\delta (S_{CS}[{\bf A}] + S^\prime _B[{\bf A}])=0$
independently of the boundary condition (\ref{BCBC}). A similar
modification is applied to the ${\bf \tilde A}$ sector.

Furthermore, the time components $A_t$ and $\tilde A_t $ play the r\^ole of 
Lagrange multipliers and can be
eliminated from the bulk action by solving the constraint equations 
$F_{r\phi} = 0$ and $\tilde F_{r\phi} = 0$ as 
$A_i = Q^{-1}_1 \partial_i Q_1$ and 
$\tilde A_i = Q^{-1}_2 \partial_{i} Q_2$,
with $i=(r, \phi)$. 
The asymptotic ${\rm AdS_3}$ behaviour (\ref {5}) implies that the
$SL(2,\real)/Z_2$ group elements $Q_1$ and $Q_2$ 
asymptotically factorize into 
$Q_1(r,\phi,t)=q_1(\phi,t) H(r)$ and 
$Q_2(r,\phi,t)= q_2(\phi,t) H(r)^{-1}$,
with 
$H(r)=diag.(\sqrt{r/\ell}, \sqrt{\ell/r})$. 
On the other hand, the components $A_t$ and $\tilde A_t$ in the boundary
action $S_B$ may be eliminated in terms of $A_\phi$, $\tilde A_\phi$,
$K_{-}$ and $\tilde K_{+}$, using the boundary conditions (\ref{BC}), which
can be re-expressed as:
\begin{equation}
\label{BC1}
A_t = \frac 1 {e^t_{-}} ( - e^\phi_{-} A_\phi + K_{-}) \qquad , \qquad 
\tilde A_t = \frac 1 {e^t_{+}} (- {e^\phi_{+}} \tilde A_\phi + \tilde K_{+})
\qquad.
\end{equation}
These relations allow to write the complete action
$S=S_{SC}[{\bf A}] +S^\prime [{\bf A}] -
S_{SC}[{\bf \tilde A}]-S^\prime [{\bf \tilde A}]$ as: 
\begin{eqnarray}
\label{30}
S =  & - \Gamma[Q_1] + & \frac 1 2 \int_{\partial \cal M}
   Tr[ \frac 1 { e_{-}^t} \, q^\prime _1 \,
 ( q_1^{-1} \partial_{-} q_1 - \, k_{-} ) ]
    \, dt \, d\phi \nonumber \\
  & + \Gamma[Q_2] - & \frac 1 2 \int_{\partial \cal M}
   Tr[ \frac 1 { e_{+}^t} \, q^\prime _2 \,
 ( q_2^{-1} \partial_{+} q_2 - \, k_{+} ) ]
    \, dt \, d\phi \qquad,
\end{eqnarray}
where $\partial _{+}$ and $\partial _{-}$ are derivatives along the
vectors $\vec e_{+}$ and $\vec e_{-}$, $q^\prime=q^{-1} \partial_{\phi}q$,
$k_{-}=HK_{-}H^{-1}$, $\tilde k_{+}=H^{-1}\tilde K_{+}H$ and
$\Gamma [Q] = {1\over{3!}} \int Tr [Q^{-1} dQ \wedge Q^{-1} dQ
\wedge Q^{-1} dQ]$.

Using the new variable $q=q_1^{-1} q_2$, 
the fields $q_1$ or $q_2$ can be eliminated 
using their equations of motion, as they only appear in  quadratic
expressions of their derivatives with
respect to the angular variable $\phi$. The resulting action becomes a
non-chiral WZW-like action containing only the field $q$:
\begin{equation}
\label{NCWZW}
S =  \Gamma[Q] - \frac 1 2 \int_{\partial \cal M}
 Tr[ q^{-1}\partial_{+} q \, q^{-1}\partial_{-} q 
    + \,2 \, \partial_{+} q  q^{-1} k_{-} 
  - \, 2 \, q^{-1} \partial_{-} q  k_{+}] 
  \, \theta \, dt \, d\phi \quad.
\end{equation}

Let us for a moment focus on the equations of $q_1^\prime$ and $q_2^\prime$
as functions of $q$: 
\begin{eqnarray}
\label{33}
q_1^\prime &=& \theta [ {e_{-}^t}  
  ( \partial_{+} q q^{-1}  - q \tilde k_{+} q^{-1} ) +
    {e^t_{+}} k_{-}]  \qquad , \\
\label{33a}
q_2^\prime &=& \theta [{e_{+}^t} 
  ( q^{-1} \partial_{-} q   + q^{-1} k_{-} q ) -
    {e^t_{-}} \tilde k_{+}]  \qquad .
\end{eqnarray}
Using the Gauss decomposition
$q = \left(\begin{array}{cc}
e^{\Phi /2} + XY e^{-\Phi/2} &X e^{-\Phi/2}\\
Y e^{-\Phi/2} &e^{-\Phi/2}
\end{array}\right)$ for $SL(2,\real)$ elements,
eqs (\ref{33}, \ref{33a}) lead to 6 equations. Four of them:  
\begin{eqnarray}
X=\frac \ell 2 \partial _ {+} \Phi & \qquad, \qquad & 
Y=\frac \ell 2 \partial _ {-} \Phi \qquad,\\
\label{XY}
\ell (\partial_{-} + \omega _{-}) X + \frac 1 2 \sigma + e^\Phi=0 & \qquad, 
\qquad & 
\ell (\partial_{+} - \omega _{+}) Y + \frac 1 2 \sigma + e^\Phi =0 \quad ,
\end{eqnarray}
determine $X$ and $Y$ as functions of $\Phi$ and combine to give:
\begin{equation}
\label{43}
\Box \Phi + \frac 8 {\ell^2} e^{\Phi} + \frac 4 {\ell^2} \sigma = 0 \qquad .
\end{equation}
This is the Liouville equation on curved background, the curvature
being given by eq. (\ref{14}).
The other equations yield relations between the energy-momentum tensor
of the Liouville field $T_{ab}$ and the metric:
\begin{eqnarray}
\stackrel {(2)} g _{ab} & = &  \frac {\ell^2} 2 T_{ab} 
- \eta _{ab} \frac {\ell ^2} {2} \stackrel {(0)}{\cal R}\\
 & = &
\frac {\ell^2} 2 [
\frac 1 2 \Phi _{;a} \Phi _{;b} -\Phi_{;ab} + 
\stackrel {(0)}{\cal R}_{ab} \Phi -\eta_{ab}
(\frac 1 4 \Phi_{;c} \Phi^{;c} +\frac 4 {\ell^2} e^\Phi +
\frac 1 2\stackrel {(0)}{\cal R} \Phi)]  \, .
\end{eqnarray}
As a consequence, 
$q$ can be expressed in terms of $\Phi$ and 
its derivatives only. Note that the elimination of the $X$ and $Y$
variables in 
the non-chiral WZW action (\ref{NCWZW}), using the constants
of motion defined by eqs (\ref{XY}), has to be performed following
 the same trick as the one that leads
to the Maupertuis action in classical mechanics when the energy
is conserved.  This yields
the equivalent action defined on the boundary only (without
any remaing bulk terms): 
\begin{equation}
\label{Liou}
S= {1 \over 2} \int_{\partial \cal M} [{1 \over 2} \stackrel {(0)} g 
\strut ^{ab}
\partial_a \Phi \partial_b \Phi - \frac 8 {\ell^2} e^{\Phi}+ 
\stackrel {(0)}{\cal R} \Phi] 
  \, \theta \, dt \, d\phi \qquad ,
\end{equation}
which is the Liouville action. Let us emphasize that the curvature term
appearing here comes directly from its definition in terms of the asymptotic
metric $\stackrel {(0)} {\bf g}$, and not through $\sigma$ as it is the case 
in eq. (\ref {43}).

To conclude, we would like to emphasize several points. First, 
the usual EH action is divergent and equal to $S_{CS}[{\bf A}]-
S_{CS}[{\bf \tilde A}]$ plus an additional term 
$\frac \ell 2 \int _{\partial {\cal M}}Tr[{\bf A} \wedge {\bf \tilde A}]$,
which is equal to half of the extrinsic curvature term usually added to
the EH action to cancel its variation \cite{MB}. 
However, owing to the AdS boundary conditions, this additional term
 does not contain any
dynamical degrees of freedom and may be dropped, thereby rendering
the resulting action finite. 
Furthermore, the developments following eq. (\ref{30}), which lead from
the non-chiral WZW action to the Liouville action (\ref{Liou}), 
are classically 
valid, but have to be re-examined in the framework of quantum mechanics.
Indeed, in quantum mechanics, the changes of variables leading to 
eq. (\ref{NCWZW}) and the
subsequent elimination of the $X$ and $Y$ variables in terms of $\Phi$
involve functional determinants that have been completely ignored.

\vspace*{0.25cm} \baselineskip=10pt{\small \noindent We acknowledge
M. Ba\~nados, K. Bautier, F. Englert and M. Henneaux for
fruitful discussions. One of us (Ph. S.) thanks the organizers of
the Journ\'ees Relativistes 99 for giving him the opportunity to present 
these results. M.~R. is Senior Research Associate at the Belgian National
Fund for Scientific Research.}

\end{document}